\documentclass[a4paper,12pt,oneside]{article} 
\usepackage{latexsym}
\usepackage{tikz}
\usetikzlibrary{arrows, shapes}
\usepackage{chngpage}
\usepackage{soulutf8} 
\usepackage{color,soul}
\usepackage{amsmath}
\usepackage{amssymb}
\usepackage{graphicx}
\usepackage{enumerate}
\usepackage{tabularx}
\usepackage{booktabs} 

\newcommand\clearrow{\global\let\rowmac\relax}
\clearrow
\usepackage[colorinlistoftodos,prependcaption,textsize=tiny]{todonotes}
\usepackage{xargs}     
\newcommandx{\unsure}[2][1=]{\todo[linecolor=red,backgroundcolor=red!25,bordercolor=red,#1]{#2}}
\usepackage[style=authoryear,bibstyle=authoryear,minnames=1,maxnames=2,maxbibnames=99,natbib,url=false,doi=false,backend=bibtex]{biblatex} 
\usepackage{dsfont}
\usepackage{xcolor}
\usepackage[]{currvita}
\usepackage{comment}
\usepackage{geometry}
\usepackage{caption}
\usepackage{subcaption}
\usepackage{fancyhdr}
\usepackage{afterpage}
\usepackage{parskip}
\usepackage[utf8]{inputenc}%
\usepackage{csquotes} 
\usepackage[official]{eurosym}
\usepackage{pdfpages}
\usepackage{hyperref}
\usepackage{lineno}
\usepackage{authblk}
\usepackage{titlesec}
\usepackage{endnotes}

\usepackage{xargs} 
\usepackage[colorinlistoftodos,prependcaption,textsize=tiny]{todonotes}
\newcommandx{\TKcomment}[2][1=]{\todo[linecolor=blue,backgroundcolor=blue!25,bordercolor=blue,#1]{TK: #2}}
\newcommandx{\JScomment}[2][1=]{\todo[linecolor=blue,backgroundcolor=green!25,bordercolor=green,#1]{JS: #2}}

\titleformat{\subsection}
{\bfseries}{\thesubsection}{1em}{}
\titleformat{\subsubsection}
{\itshape}{\thesubsubsection}{1em}{}

\titlelabel{\thetitle.\quad}
	\titlespacing\section{0pt}{0pt plus 0pt minus 0pt}{-4pt plus 0pt minus 0pt} 
	\titlespacing\subsection{0pt}{-4pt plus 0pt minus 0pt}{-4pt plus 0pt minus 0pt} 
	\titlespacing\subsubsection{0pt}{-4pt plus 0pt minus 0pt}{-4pt plus 0pt minus 0pt} 

\newcommand{\chaptermark}[1]{%
\markboth{\thechapter.\ #1}{}}
\renewcommand{\sectionmark}[1]{\markright{\thesection.\ #1}}
\begin{document}

\setlength{\belowdisplayskip}{-2pt} \setlength{\belowdisplayshortskip}{-2pt}
\setlength{\abovedisplayskip}{-15pt} \setlength{\abovedisplayshortskip}{-15pt}

\pagenumbering{gobble}

\pagestyle{fancy}
\fancyhead{}
\fancyhead[R]{\bfseries\sffamily\nouppercase  \leftmark}
\fancyhead[LO]{\bfseries\sffamily\nouppercase  \rightmark}

\renewcommand{\sectionmark}[1]{\markright{\thesection~~#1}}
\renewcommand{\chaptermark}[1]{\markboth{\if@mainmatter\chaptername\ \thechapter~~\fi#1}{}}

\newgeometry{inner=3cm,outer=3cm,tmargin=3.3cm,bmargin=3.3cm}

\title{Modelling Social Evolutionary Processes and Peer Effects in Agricultural Trade Networks: the Rubber Value Chain in Indonesia}

\author[1]{Thomas Kopp\thanks{thomas.kopp@agr.uni-goettingen.de, +49 551 394821 (Corresponding author).}}
\author[1]{Jan Salecker\thanks{jsaleck@uni-goettingen.de.}}
\affil[1]{Georg August University, Platz der Göttinger 7, 37073 Göttingen, Germany}

\date{}
\maketitle
\vspace{-1cm}
\begin{abstract}
Understanding market participants' channel choices is important to policy makers because it yields information on which channels are effective in transmitting information. 
These channel choices are the result of a recursive process of social interactions and determine the observable trading networks. They are characterized by feedback mechanisms due to peer interaction and therefore need to be understood as complex adaptive systems (CAS). 

When modeling CAS, conventional approaches like regression analyses face severe drawbacks since endogeneity is omnipresent. 
As an alternative, process-based analyses allow researchers to capture these endogenous processes and multiple feedback loops. 
This paper applies an agent-based modeling approach (ABM) to the empirical example of the Indonesian rubber trade. 
The feedback mechanisms are modeled via an innovative approach of a social matrix, which allows decisions made in a specific period to feed back into the decision processes in subsequent periods, and allows agents to systematically assign different weights to the decision parameters based on their individual characteristics. 
In the validation against the observed network, uncertainty in the found estimates, as well as under determination of the model, are dealt with via an approach of evolutionary calibration: a genetic algorithm finds the combination of parameters that maximizes the similarity between the simulated and the observed network. 

Results indicate that the sellers' channel choice decisions are mostly driven by physical distance and debt obligations, as well as peer-interaction. Within the social matrix, the most influential individuals are sellers that live close by to other traders, are active in social groups and belong to the ethnic majority in their village.
\end{abstract}

\textbf{Keywords:} agent-based modelling; inverse modelling; complex adaptive systems; networks; rubber; Indonesia; agricultural trade.

\newpage

\pagenumbering{arabic}
\setlength{\headheight}{15pt}

\section{Trade networks are complex, adaptive systems}
Understanding market participants' channel choices can provide critical insights into which networks are effective in transmitting information. This is important for policy makers who prefer making use of existing structures over introducing a new network of extension workers. 
Marketing decisions which materialize in the observed trading network are often characterized by feedback loops, especially in rural environments and small communities. 
Previous research finds evidence for 
\textquote{higher response to [\dots] conservation covenanting programs when agents are part of a local uniform matching network} \citep[p. 1]{Iftekhar2016}. This paper suggests an agent-based modelling (ABM) approach to test for the importance of social closeness and peer interaction in market participants' trading decisions. 

These decision-making processes can be referred to as \emph{complex adaptive systems} (CAS) as defined by \citet{Holland1996}, as they are influenced by multiple and complex interactions between stakeholders and the recursive nature of the decisions. \citet[p. 10]{Rammel2007} describe CAS as leading to \textquote{large macroscopic patterns which emerge out of local, small-scale interactions}. These evolutionary dynamics arise because CAS result from a) interactions amongst agents, b) interactions between agents and the environment, and c) a learning process through repetitions of these interactions. The whole system of agents therefore adapts to the environment \citep{Potgieter2005}. So every action and its respective result is fed back into the decision-making processes in future periods which is referred to by \citet[p. 10]{Rammel2007} as \textquote{co-evolutionary processes and dynamic patterns}. While the CAS approach is most often applied to biological processes, \citet{Markose2005} argues that socio-economic systems like markets may well be understood as CAS, too.

Feedback loops are a source of high-degree complexity in CAS \citep{VandenBergh2003} and do not settle to static equilibria \citep{Rammel2007}.
Conventional approaches such as regression analysis face severe drawbacks in modelling these processes since endogeneity is omnipresent \citep{Holland2006}.

Process-based approaches are better suited to study CAS since they can capture endogenous processes and multiple feedback loops. 
One promising alternative following this logic is analysis via agent-based modelling (ABM). 

A prime example of CAS at work in the field of economics is in the marketing network for natural rubber in the Jambi Province in Indonesia. 
In Jambi Province, rubber is produced predominately by smallholders whose output is distributed via a dense network of small agricultural traders to domestic processors. These are crumb rubber factories. Since traders vary in size and capacity, smaller village traders sell the rubber to larger district traders, who then sell either to the processor or to still another trader \citep{Kopp2017a}. The questions that this paper addresses are: \emph{why do agricultural traders sell to a specific buyer?} And, on an aggregated level, \emph{what are determinants of a trading network's structure?} To avoid confusion, when differentiating between selling traders and their buyers, including other traders, this paper refers simply to \textquote{buyers} and \textquote{sellers} throughout. 

To answer these questions, we model each seller's decision regarding whom to sell to as a recursive process. 
An individual seller’s initial selection of a buyer is decided through ranking all potential buyers based on various characteristics, which are weighted by global parameters. 
The decisions made by an individual seller’s peers in the previous period affect his/her decision-making process in the current period. But just how strongly do the past decisions made by other sellers affect the channel choices of the individual trader under consideration? This decision appears to be in part determined by the social closeness between the sellers. 
These effects are operationalized in our model via a so-called \emph{social matrix}, which quantifies the social closeness between each seller and his or her peers. 
The resulting matrix weights the impact of all other sellers’ decisions on each individual seller. After each iteration descriptive metrics of the predicted network are saved. This process is then repeated until the metrics converge. 
The resulting, seller-specific lists order all potential buyers according to each individual seller's propensity to engage in trade with them. An optimization algorithm is used to determine the values of the global parameters which maximize the number of correctly predicted trading links.


To summarize, this paper models the channel-choices of agricultural traders with an ABM approach. This is superior to approaches of regression analysis because of dynamic network effects introduced by feedback-loops in the agents' decisions. The model includes an innovative approach of multiplying a \emph{social matrix} with a weighting vector, allowing for heterogeneous effects based on individual characteristics. 
This enables the researcher to identify \emph{influential} individuals whose decisions are disproportionally influential. The analysis builds upon a unique dataset generated during a 2012 micro-level survey of agricultural traders in Jambi province, Indonesia.

To the best of the authors' knowledge, this is the first paper employing an ABM approach based on the theory of CAS to predict agricultural traders' channel choices. The inclusion of a social matrix in traders' network analysis is an innovative approach as well. 
The methodology developed in this paper can be employed in other areas where the identification of efficient channels for transmitting information is desired.

This paper is structured as follows: the literature review section gives an overview of ABM based approaches used in the agricultural economics literature so far. Section three presents our ABM RUBNET. The complete model description (ODD protocol: overview, design concepts, and details) is included in the appendix. 
Section four presents and discusses the results, and section five provides a conclusion.


\section{ABM and CAS in the (agricultural) economics literature}
While agent-based modelling is increasingly being applied in agricultural economics literature, virtually no empirical work has been undertaken to model marketing decisions at the micro level. The majority of studies model production decisions at the farm level, such as investment decisions \citep{Feil2013,Resende-Filho2008}, adaption of new technologies \citep{Schreinemachers2009}, participation in certification schemes \citep{Latynskiy2017}, farm-level climate change adaptation \citep{Troost2014}, or breeders' responses to price shocks \citep{Zhang2010}. 
\citet{Latynskiy2017} model decision making within coffee farmers cooperatives to understand processes of collective action in voluntary sustainability certifications.
On a similar track, \citet{Iftekhar2016} model the participation of farmers in conservation programs subject to constraints on land-use and existing social networks. 
\citet{Boyer2013} analyse processors' market power by including the US Livestock Mandatory Price Reporting Act into an auction-based ABM.

In a broader economic application, \citet{Klos2001} analyse transaction costs with an ABM to explicitly account for often-ignored characteristics of transactions like mutual trust and heterogeneity of agents, and challenge the assumption of efficient outcomes.
\citet{Alfarano2009} use an ABM approach to generate evidence on macro level outcomes from the behavior of a multitude of diverse agents on a micro level on financial markets. 
\citet[p. 1182]{Zhang2010} argue that agent-based computational economics (ACE) \textquote{can be used to study problems with behavioral assumptions that are too difficult to analyze with mathematical methods. ACE is more economical and time efficient compared with experiments with human subjects (e.g. \citealp{Ward1999}) and is more controllable.}.

In the CAS literature, \citet{Markose2005} provides an extensive overview of ABM approaches in analysing CAS in economics. These are required in situations that deviate from the basic assumptions typically made in economics, for example when understanding processes such as \textquote{innovation, competitive co-evolution, persistent heterogeneity, increasing returns, the error-driven processes behind market equilibria, herding, crashes and extreme events such as in the business cycle or in stock markets.} \citep[p. 159]{Markose2005}. 
\citet{Butler2016} reviews the literature on applications of complex systems to agricultural economics. He generally concludes that nowadays (agricultural) markets are too complicated for regression analysis, which assumes the emergence of equilibria, because markets need to be understood as complex systems \textquote{in which economic agents [\dots] continually adjust and react to market behavior of others} \citep[p. 2]{Butler2016}. Reasons for this perceived complexity include feedback loops and neighbor effects, which are likely to occur in the seller’s decision-making process. 
\citet{Chen2001} analyse the behavior of traders on an artificial stock-market with an ABM. They find that while traders may behave as if they do not believe in the efficient market hypothesis, their aggregate behavior results in an efficient capital market.

When studying trading networks, we are observing a highly dynamic system in which each seller's decisions have an influence on each of his or her neighbors' decisions, which might then in turn influence other neighboring traders. Since these effects go back and forth, endogeneity is omnipresent, which prevents the use of conventional regression analysis. 
And whenever observations are only available for one point in time it is difficult to assess \emph{how knowledge spreads} between stakeholders. The ABM approach is a way of circumventing the endogeneity problem \citep{Zhang2010}. We therefore employ an agent-based, pattern-oriented modeling approach to generate a hypothetical outcome under certain assumptions (represented as model parameters, \cite{Grimm2005a}). We predict trading connections of model agents based on sets of parameter values and compare the emerging network to the observed network. Global parameters are then changed systematically in order to maximize the level of resemblance between the simulated and empirical trading network.

\section{Application to the rubber value chain in \mbox{Indonesia}}
Value chains in rural areas of less developed countries tend to rely on agricultural traders and middlemen. These key agents offer crucial services, such as the transportation of farm output, the reduction of information asymmetries (for example on prices), and the lending of credit to farmers. 
The traders we observed in the rubber market value chain in Jambi Province, Indonesia, provide all of these services.  

\begin{figure}[h t b]
	\protect\caption{Trader Network in Jambi.\label{f.network}}
	\centering
	\includegraphics[width=100mm]{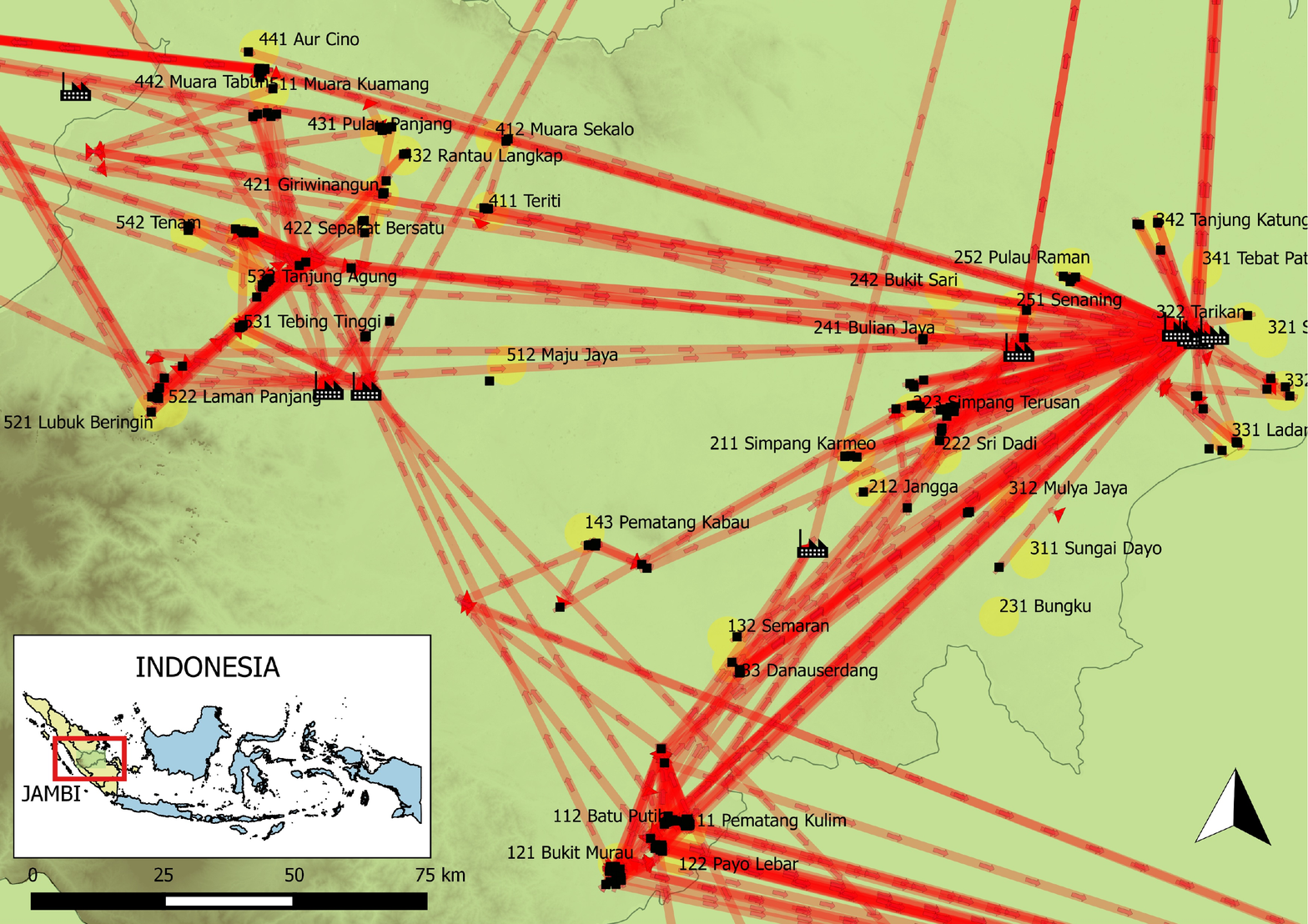}
	\caption*{Source: Own production, based on original survey data, collected in a representative survey with rubber traders in 40 villages in the Jambi Province (Sumatra, Indonesia) as well as with the downstream traders that the initial respondents named as their buyers. Further information on the sample can be found in \citep{Kopp2017a}. Borders of Jambi and Sumatra from Center for International Forestry Research, surface of Jambi from NASA/EOSDIS.}
\end{figure}

Unlike many other agricultural products, Rubber is not perishable. This enables the formation of long value chains consisting of many stakeholders, resulting in extended networks between traders, as can be seen in Figure \ref{f.network}.

\subsection{Model description}

The overall aim of our study is to reconstruct the channel choice behavior of selling agents within an empirical context. 
%
%
So the specific purpose of RUBNET is to understand channel choice behavior of rubber traders in Jambi, Sumatra, Indonesia. 
Testable hypotheses are:
	\begin{itemize}
	\item $H_1$: Only distance matters in the marketing decision.
	\item $H_2$: Sellers sell to the buyer that offers the highest price.
	\item $H_3$: Neighboring sellers influence each other.
	\item $H_4$: If a seller has a credit with a specific buyer, he/she cannot sell to another one.
	\end{itemize}

We run an optimization scenario by using genetic algorithms that vary weight parameters in order to maximize the proportion of correctly predicted trading links \citep{Kumar2010}.

RUBNET is implemented in NetLogo version 6.0.3 \citep{Wilensky1999}. 
The detailed ODD (Overview, Design concepts, Details) protocol for describing individual-based models \citep{Grimm2006a,Grimm2010} is provided in the \hyperlink{appendix.ODD}{appendix}. 

\subsubsection{Behavioral assumptions}

Traders are assumed to make decisions, which are mainly influenced by revenue and transaction costs of selling, in order to maximize their profits. 
Revenue depends on the price they receive from their buyer. Transaction costs depend on the location relative to the seller and other unobserved characteristics of the buyer. The selection of a buyer therefore crucially affects the seller's profits. Personal relationships between stakeholders are assumed to play a central role in the decision-making process as well. They include interactions between the trader under consideration with a) potential buyers and b) with his or her peers, i.e. neighboring sellers. The relationship between the seller and potential buyers is characterized by a) stable characteristics such as the physical distance and b) time-variant characteristics such as the amount of credit the seller has taken from the buyer in previous periods. The relations between neighboring sellers include spill over and learning effects and therefore generate feedback mechanisms. \textquote{Neighborhood} is defined widely, i.e. along a number of dimensions such as physical proximity, ethnicity, and similarity in level of education. 
The causal chain of an agent's decision making is displayed in Figure \ref{FigureCausalChain}.

\begin{figure*}[h t b]
\protect\caption{Causal chain of decision making process 
\label{FigureCausalChain}}
\begin{center}
\begin{tikzpicture}
[> = latex', auto,
block/.style ={rectangle,draw=black, thick,
align=flush center}, 
]
\matrix [column sep=5mm,row sep=7mm]
{
\node [block,rounded corners=0.3cm] (box1) {\textbf{Stable determinants}\\
Characteristics of buyer:\\
$\bullet$ Location\\
$\bullet$ Unobservable characteristics
};
&
\node [block,rounded corners=0.3cm] (box2) {\textbf{Variable}\\ \textbf{determinants I}\\
Seller-buyer relation:\\
$\bullet$ Debt of seller with\\specific buyer
};
&
\node [block,rounded corners=0.3cm] (box3) {\textbf{Variable}\\ \textbf{determinants II}\\
Network effects:\\
$\bullet$ Decisions of peers
};
\\
&
\node [block] (box4) {\textbf{Seller's decision}\\
Which buyer to sell to?};
&
\\
};
\begin{scope}[->, very thick]
\draw (box1.south) |- (box4.west);
\draw (box2.south) -- (box4.north);
\draw ([xshift=2mm]box3.south) |- ([yshift=-2mm]box4.east);
\draw ([yshift=2mm]box4.east) -| ([xshift=-2mm]box3.south) node[midway,below] {\textbf{*}};
\end{scope}

\end{tikzpicture}
\end{center}
\caption*{Source: own design\\ 
*: This feedback-loop continues until a temporary equilibrium is reached, i.e. until the network is identical in two subsequent periods.}
\end{figure*}

\subsubsection{Entities, state variables, and scales}

\begin{figure}[h t b]
	\protect\caption{Graphical representation of all agents and links \label{model}}
	\centering
	\includegraphics[width=110mm]{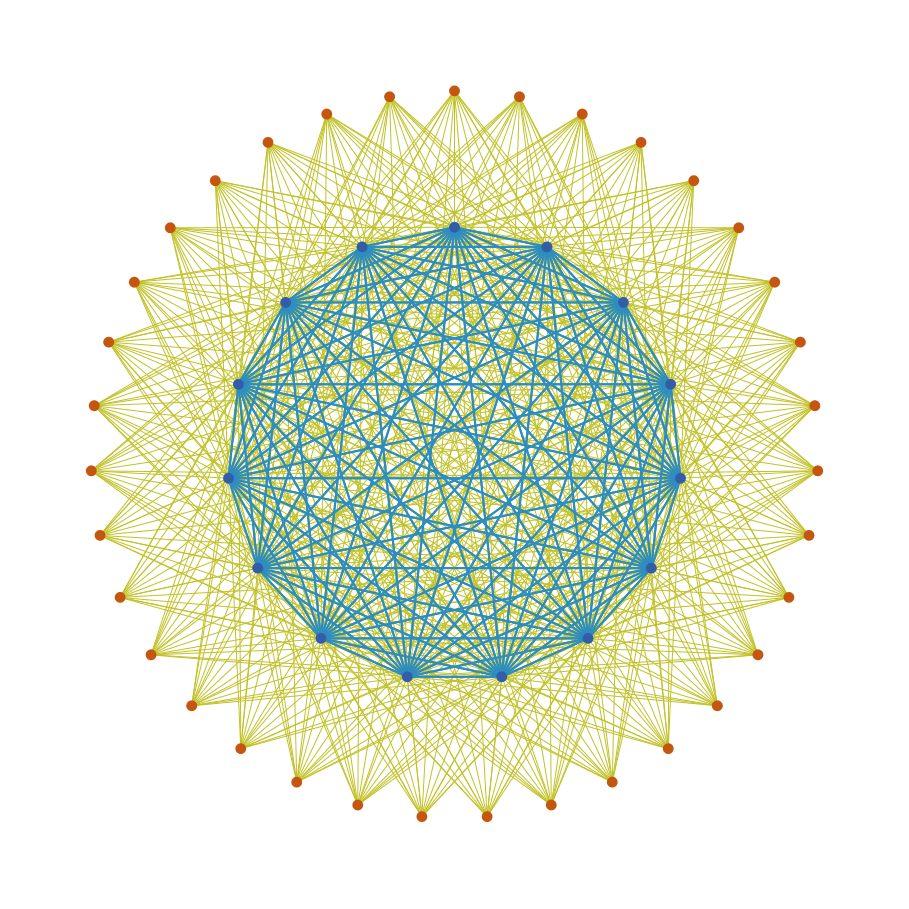}
	\caption*{Red dots indicate buying agents (h1s), blue dots indicate selling agents (h1s). Yellow links represent trading connections between selling (h0s) and buying (h1s) agents. Blue lines represent social connections between selling agents (h0s). Note that there are no yellow connections between blue dots, i.e. the sellers are not modelled to be able to sell to each other, which is confirmed by the observed data. The total number of agents has been reduced for this illustration.}
\end{figure}

The main entities of RUBNET are the trading agents, represented by network nodes, and their potential interactions, represented by network links. Trading agents are grouped into two categories: selling agents (h0s) and buying agents (h1s). All selling agents are interconnected via social links, which represent the potential social interactions of these agents. Each selling agent is further connected to every buying agent via trading links, which represent potential trading connections (Figure \ref{model}). All these links are created during model initialization. However, not all links are considered active, which is indicated by a state variable of the links. Inactive links still exist in the model but are not shown in the visual output or considered for output measurements.
Each agent and link is characterized by a set of state variables (see Tables \ref{t.var.h0s}, \ref{t.var.h1s}, and \ref{t.var.links} in the appendix). Most of the agent and link variables, such as locations, social characteristics, and trading information are derived from empirical data. Thus, our model agents represent the original survey data distribution of agents and their characteristics.
RUBNET is spatially-explicit but does not utilize a discrete cell lattice with a specified grain and extent. Spatial distances between agents were calculated using geographical coordinates and are stored as a state variable of the links that connect the model agents. This is common practice for network models and allows to arrange agents and connections freely in space, without changing the spatial relations of the agents.

Analogous to Figure \ref{FigureCausalChain}, \emph{three sets of variables} enter the simulation: first are \emph{stable determinants}, which include characteristics of buyer-seller-pairs that are constant over time. The second set are \emph{variable determinants (1)}. These include variable characteristics of buyer-seller-pairs. Set three consists of \emph{variable determinants (2)}, which represent the network effects. More information is provided in the paragraph \textquote{Basic principles} in appendix Section \ref{BasicPrinciples}.

The selection of buying agents by the selling agents is performed by calculating scores for different criteria, such as distance or price. These criteria are weighted for score calculation. So-called \emph{global weights}, that are applied for all selling agents, are provided as global parameters (see Table \ref{t.var.global} in the appendix). Additionally, each selling household has a set of individual weight preferences that are estimated from the selling agents' individual properties (see Table \ref{t.var.h0s} in the appendix). These preferences are used to modify the global weights for final trading link score calculation.

\subsubsection{Process overview and scheduling}

First, the model is initialized (see Figure \ref{f.flowchart}, for details on initialization see appendix Section \ref{sec_init}). Based on empirical data from an extensive survey of rubber traders, selling and buying model agents with specific agent IDs are created. Directed social links are created from each selling agent to each other selling agent and directed trading links are created from each selling agent to each buying agent. Geographical distances between agents are loaded from a distance matrix that provides Euclidean distances for all agent ID pairs. These distances are stored as link state variables. In order to initialize the trading related state variables of links and agents, the trading data set from the survey is loaded into the model. The data provides actual trading connections, including variables such as amount of rubber traded, selling and buying prices, debts between selling and buying agents, and social characteristics of agents.
After all agents and links are initialized, preliminary scores are calculated for each outgoing trading link of each selling agent. First, each selling agent calculates sub-scores for all outgoing trading links for the criteria price, distance, and debts. These individual scores are stored as an individual characteristic of the trading links. Then a preliminary final score is calculated using the global weights variables and the individual weight preferences.

The model execution covers four main procedures: determination of social interactions, calculation of final trading scores, selection of active trading connections, and calculation of output metrics (see Figure \ref{f.flowchart}). To determine social interactions, a social influence score is calculated for each social link, based on the properties of the two connected selling agents (for details see appendix Section \ref{sec_submodel_social}). Depending on the parameter n\_social each selling agent selects the most influential incoming social links to be active. These active social links are considered the social network in terms of interactions between selling agents. Afterwards, final scores for each trading link are calculated. In contrast to the preliminary scores, these final scores also incorporate the trading decisions of socially connected selling agents. Finally, active trading connections are selected based on the final trading scores, and output metrics of the emerging trading network are calculated.

\begin{figure}[h t b]
	\protect\caption{Flowchart of model processes. \label{f.flowchart}}
\begin{center}
	\includegraphics[width=80mm]{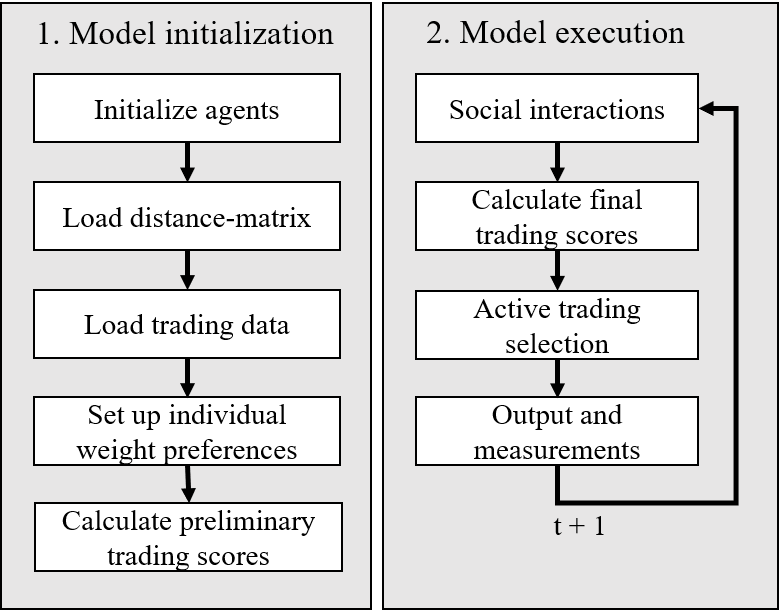}
\end{center}
	\caption*{Main procedures during model initialization and model execution.}
\end{figure}


\subsection{Scenarios}

We utilize the parameterized RUBNET model to simulate five scenarios by varying the initial properties of traders in the network. 
The indicators of interest are a) the mean prices paid in the resulting network, b) the mean length of active trading links, and c) the number of buyers being chosen, which corresponds to the size of network components. (a) is of interest because a possible price increase realized by sellers would be, even if only partial, passed on to small scale farmers, increasing province-wide standards of living. The distance between buyers and sellers (b) is decisive for the transaction costs to be incurred per sales instance. The number of network components (c) is interesting because it measures market structure, which is found by \citet{Kopp2017a} to be a determinant of market power: buyers located in villages with fewer competitors are more likely to exercise oligopsonistic market power than in others, \emph{ceteris paribus}. 
The following scenarios are evaluated:
\begin{itemize}
\item A1: All debts between sellers and buyers are halved (A2: set to zero), representing a policy to increase the share of credit that is taken from formal lending institutions.
\item B1: Increase the sellers' education from 1 and 2 to 3, i.e. ensuring primary education for all. B2: Increase the education to the highest level in each village, i.e. ensure that everybody visits the schools that already exist. 
\item C: Homogenize transport capacity (set the capacity of the lower half of the population to the mean value) to predict the results of a policy that subsidizes transportation. 
\end{itemize}

\section{Results}

\subsection{Optimization}

We implemented genetic algorithms using the R-package \textit{genalg}, version 0.2.0 \citep{Willighagen2015} and the openMole software. All global weight parameters were varied within the interval [0-100] in order to maximize the proportion of correctly predicted trading links. The best parameter combination resulted in a proportion of 49\% correctly predicted trading links 
(see Table \ref{t.optimization}). 

\begin{table}[!htbp]
	\caption{Results of the optimization experiment using genetic algorithms. \label{t.optimization}}
	\begin{center}
	\begin{tabular}[l]{p{4.5cm} p{3cm} p{3cm}}	
		\toprule
		parameter        & $value.complete$&$value.reduced$\\
        \midrule
        $n\_social	$				& $05.12$   & $01.61$\\
		$w\_price	$				& $00.23$   & $03.30$ \\
        $w\_dist    $       		& $01.74$   & $12.12$ \\
		$w\_debts   $        		& $94.48$   & $64.07$ \\
		$w\_social  $        		& $03.55$   & $20.52$ \\
        \midrule
		$w\_social\_education   $   & $31.34$   & $05.96$ \\
		$w\_social\_ethnicity   $   & $10.32$   & $09.12$ \\
		$w\_social\_activegroup $   & $14.90$   & $09.91$ \\
		$w\_social\_prestigious\_job$&$02.28$   & $00.01$ \\
		$w\_social\_proximity   $   & $41.16$   & $75.01$ \\
		\bottomrule
	\end{tabular}
	\end{center}
	\caption*{$value.complete$ refers to the optimization based on the full sample while the right column -- $value.reduced$ -- refers to results based on the reduced sample (see paragraphs \textquote{Input Data} in appendix Section \ref{sec_init}). \\
Start values for each parameter were chosen randomly within the weight interval [0-100]. The population size of the genetic algorithm was 100 and we simulated 1000 iterations. The algorithm changes weights systematically in order to maximize the proportion of correctly predicted trading links. The best-found solution resulted in 49\% correctly predicted trading links.
}
\end{table}

The results for the main weights based on the reduced sample are not fundamentally different from the ones based on the full dataset. This ensures that the reduction of the sample does not bias the aggregate results. Only the parameter of the social matrix increases substantially, as expected. The following interpretation is based on the results from the reduced sample.\footnote{Elaborations on the sub-sampling procedure are provided in the paragraphs \textquote{Input Data} in appendix Section \ref{sec_init}.}

The final weights of this simulation indicate a high weight for debts and an intermediate weight for social interactions, whereas the weights for the distance and price criteria are rather small. Prices and the distance to a potential buyer do not seem to have a strong effect for choosing trading links. The social matrix has a substantial effect. The sub-weights of the social matrix are high for proximity (same village) and intermediate for ethnicity, education, and social group membership, while having a prestigious job does not make a seller disproportionally influential.

\begin{figure}[!htbp]
	\protect\caption{null models \label{f.results.nullmodels}}
\begin{center}
	\includegraphics[width=\columnwidth] {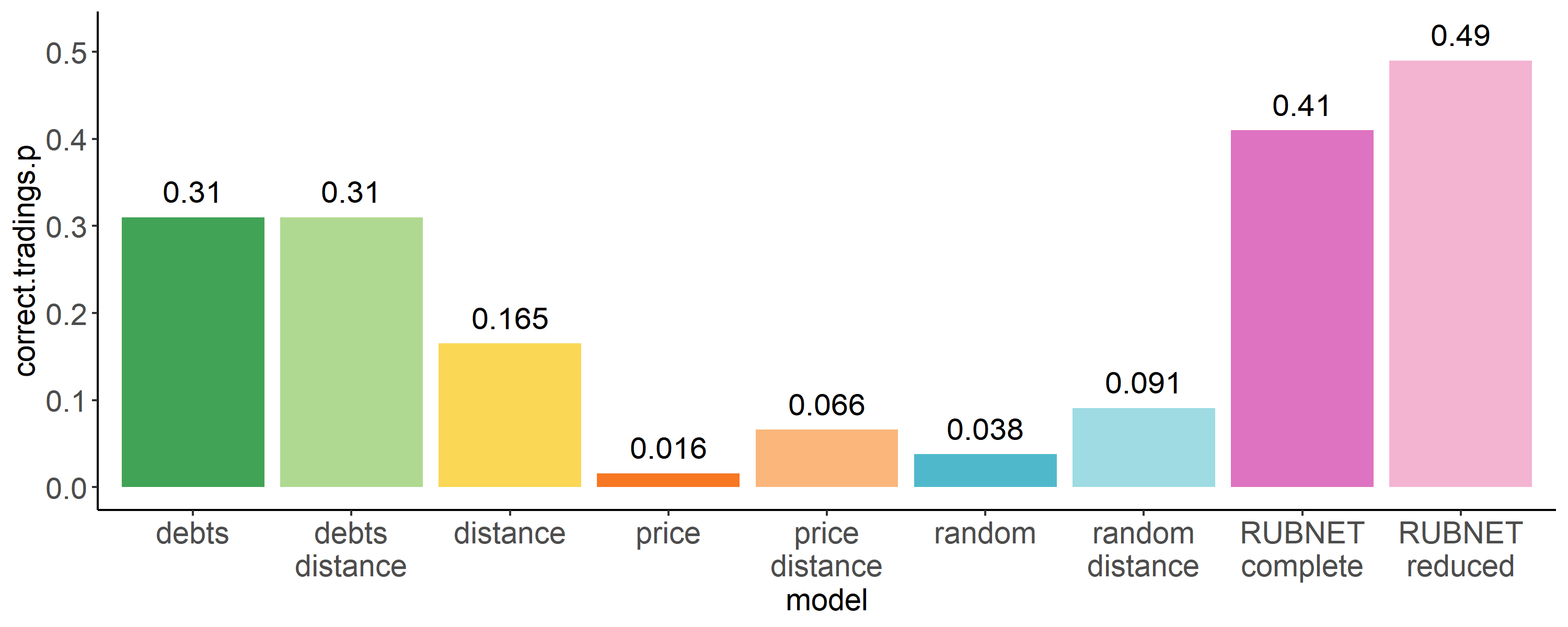}
\end{center}
\caption*{The value of each bar indicates the success rate of the parametrization found by the optimization algorithm. In the null models that include distance (debts-distance, price-distance, and random-distance) only the 25\% shortest links are allowed to be set to active.}
\end{figure}

We also compared our predictions to several null models (see Figure \ref{f.results.nullmodels}). The combined weight approach resulted in the best prediction 49\%, followed by the null model that incorporates debts only (31\%). In the random null model, which is used as a benchmark, each selling agent was assigned links to random buying agents. It predicted 3.8\% of all links correctly. The null model which only incorporates prices resulted in the worst prediction quality (1.6\%). Initially this may seem surprising, but it follows the inherent logic of the model: if only price determines the decision, the model (falsely) predicts all selling agents to decide for the same buyer. This is below the success probability of the random model. 

\begin{figure}[!htbp]
\centering
\begin{minipage}[t]{.45\textwidth}
	\centering
	\protect\caption{Trading network derived from simulations.\label{f.nw.predicted}}
	\includegraphics[trim=3.2cm 2cm 2cm 2cm , clip, width=\columnwidth]{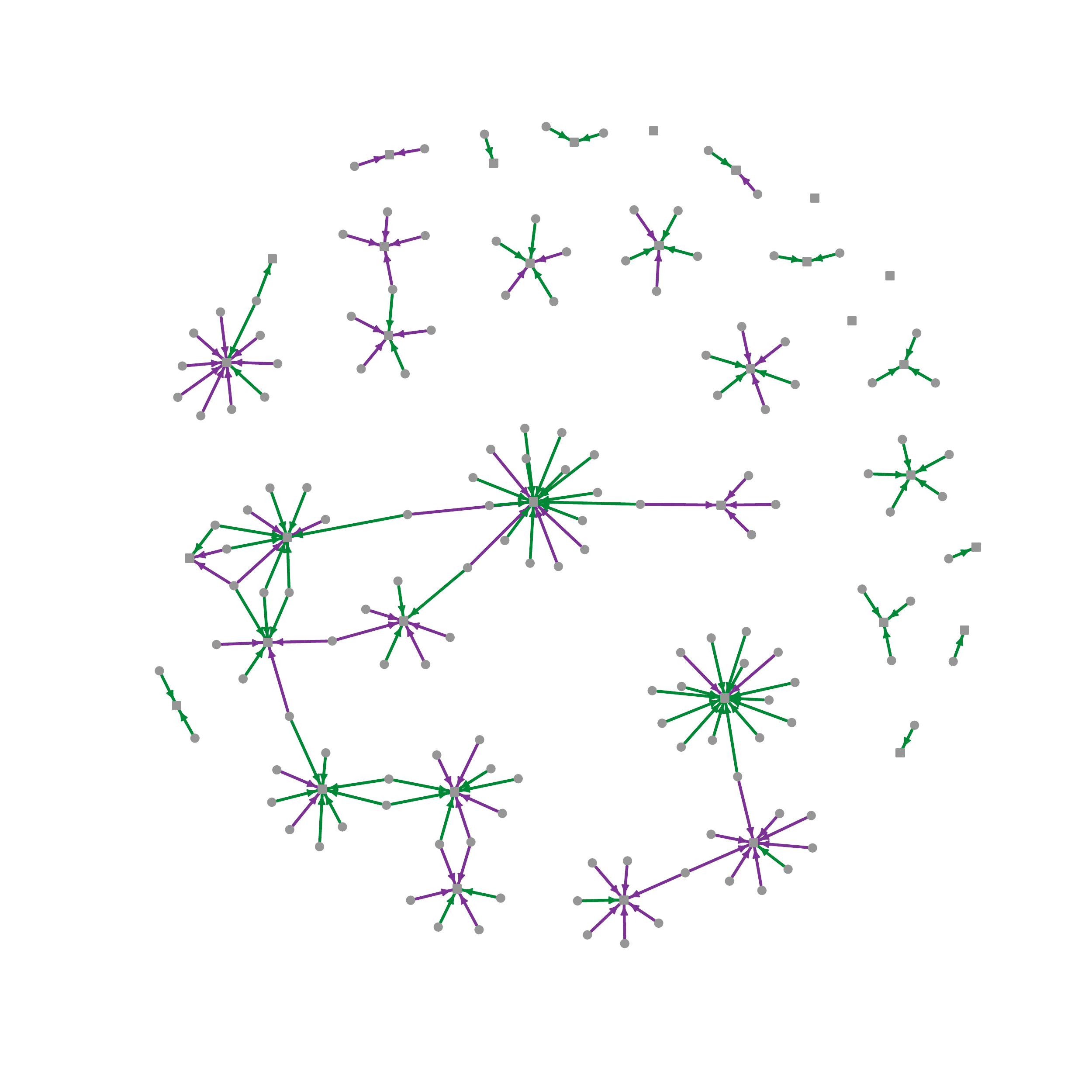} 
	\caption*{
Grey circles indicate selling agents, grey squares indicate buying agents.\\Green links indicate predicted trading connections that match empirical data. Magenta links indicate predited trading connections that do not match empirical data.\\
}
\end{minipage}%
\hspace{.05\textwidth}
\begin{minipage}[t]{.45\textwidth}
	\centering
	\caption{Trading network derived from empirical data.\label{f.nw.data}}
	\includegraphics[trim=3.2cm 2cm 2cm 2cm , clip, width=\columnwidth]{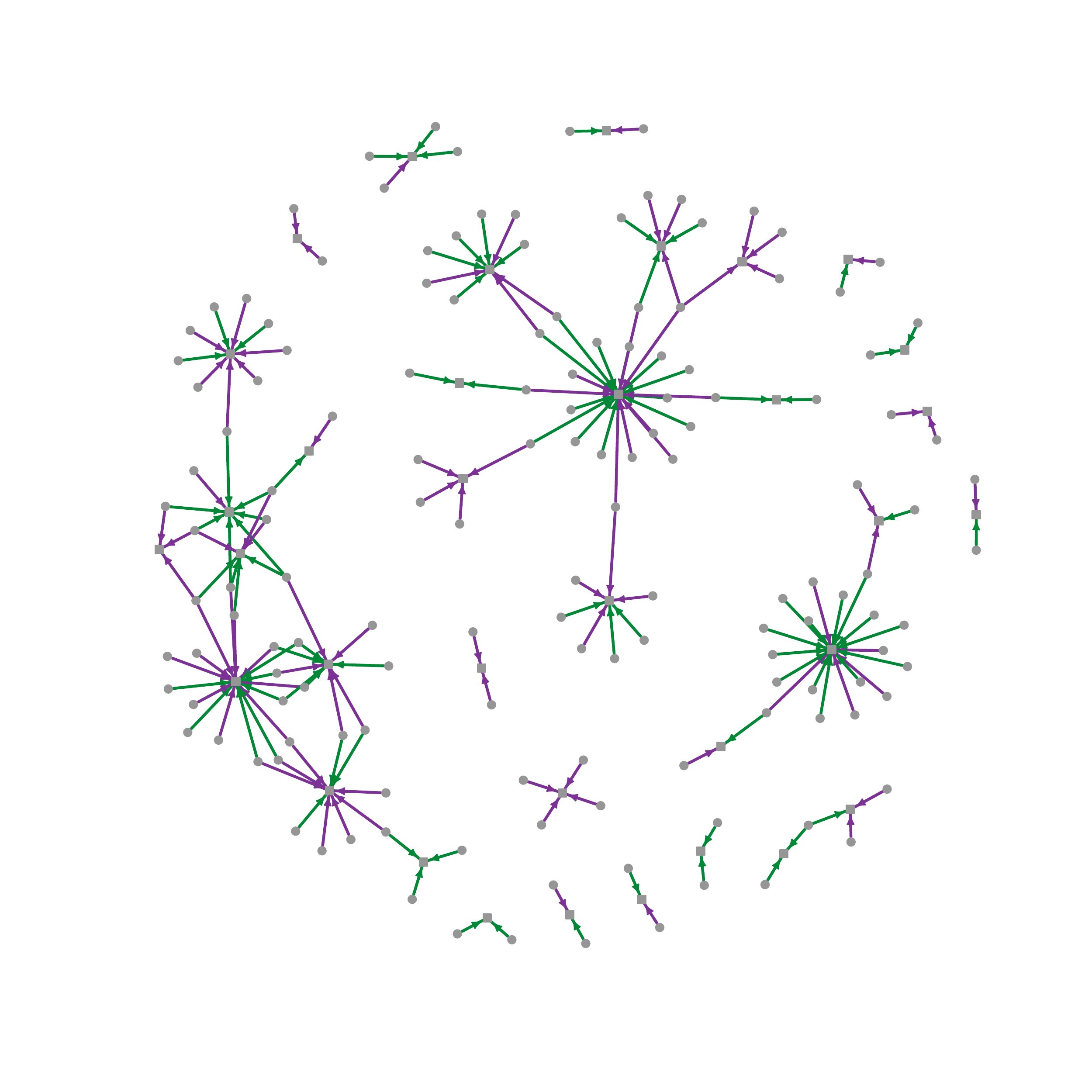}
	\caption*{Green links indicate trading connections that are present within empirical data and predicted by the model. Magenta links indicate trading connections that appear within empirical data but have not been predicted by the model. }
\end{minipage}%
\end{figure}

The resulting trading network of this optimized solution is visualized and compared to the original empirical trading network in Figures \ref{f.nw.predicted} and \ref{f.nw.data}. Standard graph metrics, such as number of components and component size, are used to quantify the structural differences. Compared to the empirical trading network, the overall number of components is higher in the predicted network (model 23; data 17). The higher number of components is a result of the four potential buying agents that have not been chosen by any selling agent. If we only consider the active trading network and ignore those agents which are not connected by any trading links, our predicted trading network is more compact and has a number of components similar to the empirical trading network (model 19; data 17). Because both networks have the same total number of active trading links, the higher number of components in our predicted network results in lower component sizes compared to the empirical trading network (mean component size model 7.8; data 10.8). 

\begin{figure}[!htbp]
	\caption{Characteristics of sellers whose channel choices RUBNET predicts well/badly.\label{f.results.nw.social}} 
	\begin{center}
	\includegraphics[width=0.9\columnwidth]{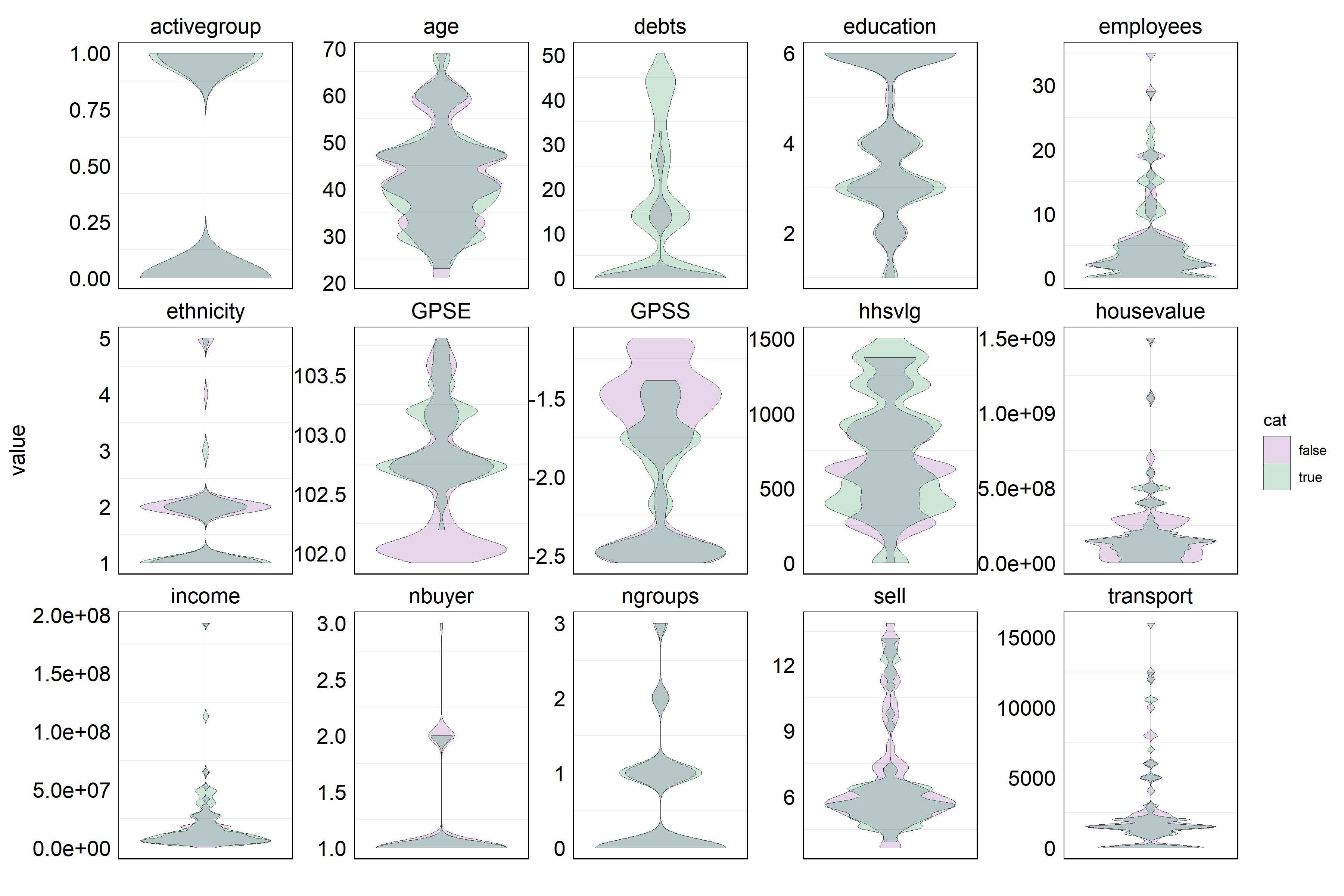}
	\end{center}
\caption*{The green shaded areas indicate the number of sellers within certain variable categories whose channel choices can be predicted well, while purple indicates the number of sellers whose purchasing decisions are poorly predictable. If the two areas overlap perfectly then the number of wrongly and correctly predicted links is equal. For example, the decisions of sellers in the highest age group are not predicted very well by RUBNET, while younger people's choices are predicted correctly more often than not.}

\end{figure}

The agent-based modeling approach allows us to investigate the individual characteristics of selling traders. Most of the distributions of social characteristics did not differ strongly between sellers whose decisions were predicted correctly and agents with only non-matching predictions (Figure \ref{f.results.nw.social}). 
For some characteristics, however, the insights deserve mentioning: for the case of debts, RUBNET correctly predicts all trading connections of buyer-seller pairs with a long history of credit transactions. This success rate decreases to 50\% when credit is zero. The selling decisions among ethnicity 3 (Sundanese) can be predicted with 100\% accuracy, while those of ethnicity 4 (Melayan) cannot be reliably predicted at all. It needs to be mentioned, though, that these two ethnic groups are especially small (5 and 3 representatives, respectively, see Table \ref{t.descriptives.categorial}), so results might suffer from limited sub-sample bias. The decisions of autochthonous traders (group 2) could be predicted with below-average accuracy. 
The decisions of older traders (above 55 years) are more difficult to predict than most other age groups. Also, the choices of sellers located in the North-West of the province can be predicted particularly poorly. These are the respondents located in Tebo and Bungo districts (see Figure \ref{f.network}). When differentiating along all other dimensions, the share of correctly and wrongly predicted links appear roughly symmetrical. 

\subsection{Discussion of optimization results}

The target of developing and implementing RUBNET, an individual-based simulation model, was to test four main hypotheses regarding rubber traders channel choice:

$H_1$: Only distance matters; $H_2$: Sellers sell to the buyer that offers the highest price; $H_3$: Neighboring sellers influence each other (i.e. coefficient of network weighting matrix $\neq 0$); $H_4$: If a seller has a credit with a specific buyer he/she cannot sell to another one (this justifies exceptions from other rules).

$H_1$ can be rejected. While distance is an important factor in determining the observed network between traders, it is not the only one. $H_2$ can be rejected, too. The null model that only includes price performs worse than all other pure models. 
It is surprising that the prices paid by potential buyers plays such a minor role. These are much less important than the distance and the indebtedness between a seller and all potential buyers. This supports the results of \citet{Kopp2017a} who find on farmers level that sellers often cannot make their own free decisions on whom to sell to if they are indebted.
$H_3$ can be confirmed. The best models in terms of predictive quality are the ones that attribute a value substantially above zero to the weight of the social matrix. $H_4$ can be confirmed, too. Being indebted with one seller is one of the two key variables affecting the predictive quality of RUBNET.

Although RUBNET cannot explain all patterns in the empirical trading network, it does identify the overall importance of social interactions and other measured variables on trading channel choice well. The individual-based approach allows us to investigate the characteristics of agents whose decisions could be predicted accurately by RUBNET and compare them to the characteristics of agents whose decisions could not be predicted well. The analysis reveals that the selling decisions of large sellers, characterized by a large number of employees, large transport capacity, and high selling volume, were particularly poorly predictable. This could possibly be explained by large sellers assigning systematically different weights to the decision parameters. If transport capacity is high, the physical distance to potential buyers may play a relatively smaller role than for sellers with less capacity for transporting goods. Also, larger sellers are generally indebted less often. A third explanation is that larger sellers more often sell to factories -- a channel characterized by high path-dependency due to switching costs \citep{Kopp2017b}. The analysis also indicated weak predictability of sellers located in the Tebo and Bungo districts. These two regions lie along the banks of the longest river in Sumatra, the Batanghari, and its tributaries, which in combination with a lack of bridges and other infrastructure shortcomings effectively splits the area in half. This is not accounted for in RUBNET due to the lack of geospatial data on the water system in the area. The high predictability of traders with high credit can be explained by the sheer importance of that variable in RUBNET. 

\subsection{Scenarios} 

\begin{figure}[h t b]
	\protect\caption{Results of scenarios.\label{f.scenarios}}
	\centering
	\includegraphics[width=1.0\columnwidth]{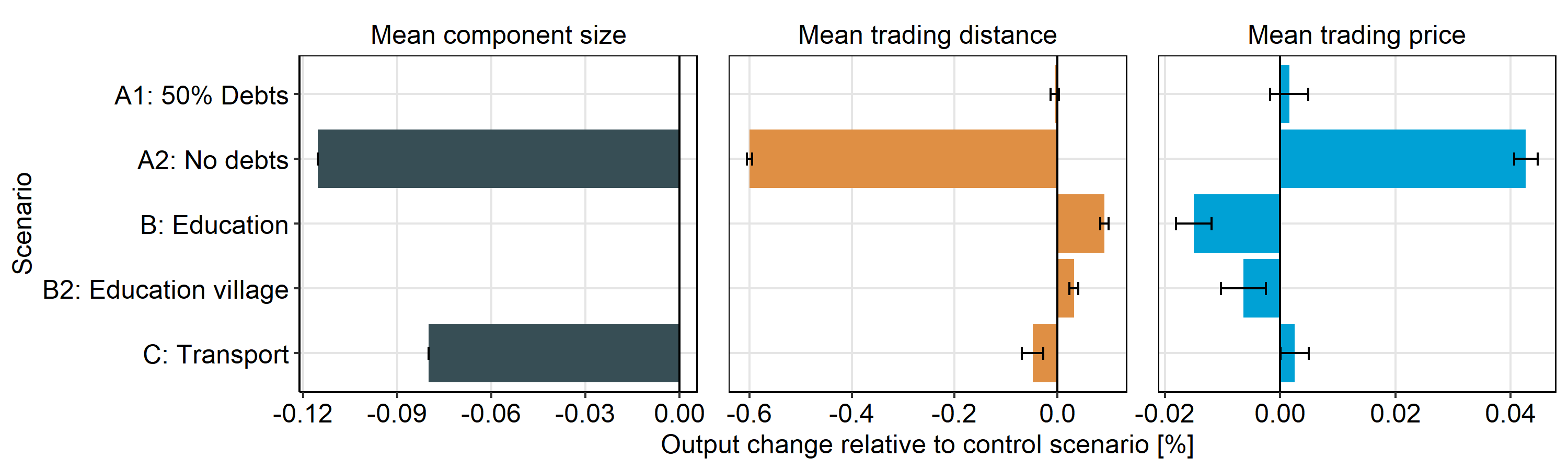}
	\caption*{Each line of bars captures the outcomes of one scenario and the three sets of bars represent one indicator each. The whiskers display the standard deviation between 20 repetitions with different random seeds.}
\end{figure}

The results of the scenarios are displayed in figure \ref{f.scenarios}. 
Since the debts variable is the biggest determinant of the sellers' choice, modifications to its values result in the biggest changes. While halving every seller's debts has virtually no effect, the (albeit unrealistic) scenario of a complete abolition of debts leads to an increase in price and a reduction in the distances between sellers and buyers. The reason for these results is that, holding all else constant, the sellers are less restricted in their choices and therefore decide to sell to a closer buyer at a higher price. 
This price increase has a positive effect for the primary producers, as at least part of the increase can be assumed to be passed on. The reduction of distance is ambiguous. While shorter travel distances between buyer and seller are likely to reduce transaction costs, it is also a key decision factor for the number of stakeholders required to be in the trading chain as a whole, as shortening transaction distances may increase the number of traders through whose hands the product passes. \citet{Kopp2017a} find that their number is positively correlated with traders' market power towards farmers, so lengthening the chain by reducing the physical distance of the predicted links will increase the market power that traders can exercise over smallholders. The number of clusters reduces (i.e. the mean cluster size increases) what can be understood as an increasingly concentrated market structure, an indicator of more efficient markets. 
This shows that policies which reduce sellers' dependence on credits from buyers will increase the smallholders' income situation. 

An increase in transport capacity has a similar effect as the abolition of all debt in that it increases average prices and reduces both trading distances and the component size. Effects are much smaller, though, than those of a full debt reduction. This means that policy measures which increase sellers' transport capacities can emulate the effects of the much less realistic scenario of abolishing all debt. 

An alignment in the sellers' education levels causes the opposite effects as the other two scenarios: prices decrease while the mean trading distance increases. The reason is that the alignment in education levels increases the number of other sellers that influence the seller under consideration, which results in the seller being informed about more attractive buyers further away. Since the modelling results indicate that price is not a key determinant of the decision, the seller is willing to sell to buyers considered to be more attractive but further away, even at the cost of a price disadvantage. 

\subsection{Outlook} 
With our modeling approach we can explain up to 49\% of the trading network connections. In order to increase the predictive quality of RUBNET, several model extensions could be implemented. Currently, RUBNET uses Euclidean distances between agents to estimate their proximity. However, in reality, road connections and barriers such as large rivers or mountain ranges affect agents' proximity and are not considered yet due to lacking road network data. In order to enhance the predictability of social-adaptive processes, more detailed data on social characteristics of trading agents can be implemented into RUBNET easily, once such data is available. 

At its current state, RUBNET does not incorporate any temporal dimension.
Traders' channel choices are rather estimated for one point in time, based on empirical data. However, our model already provides detailed insight into the economic behavior of trading agents, which is an essential step not only for future extensions of RUBNET, but also for development of other agent-based models featuring temporal dynamics, such as the economic-ecologic land-use change model EFForTS-ABM \citep{Dislich2015}. Explicit consideration of spatially explicit trading networks in a dynamic ecological-economic model will enable even more sophisticated network analyses like trading network resilience under various scenarios covering price dynamics, consolidation processes, or policies.

One huge advantage of the pattern-oriented modelling approach is that we are able to fit the model parameters based on empirical data \citep{Grimm2005a}. The resulting parameterized model can then be used to predict similar trading networks based on characteristics of traders, without any knowledge of the existing trading connections between trading agents. The quality of such predictions can be assessed and validated by using another empirical dataset which might be available in the future. 


\section{Conclusion}
This is the first paper to model the channel choice behaviors of agricultural traders with an agent-based model. This approach is motivated by the trading network of choice being shaped by social interactions and feedback loops which hinder conventional regression analysis through problems of excessive endogeneity. 
Although RUBNET, the ABM developed in this work, cannot explain all patterns in the empirical trading network, it does succeed in sufficiently identifying the overall impact of social interactions and other measured variables on traders' channel choice. 

Results show that physical distance between a seller and potential buyers, as well as the social closeness among sellers are the two most important factors when simulating traders' channel choices. Within the social matrix sub-model, the key drivers of predictive quality are physical proximity to the seller's peers, ethnicity, and the involvement in social groups.

Based on these results, several scenarios were simulated. Results indicate that a policy reducing sellers' dependence on credit from their buyers might increase the smallholders' income situation. Policies that increase education levels and transport capacities have little effect on market outcomes.

Policy-makers who want to efficiently disseminate information upstream along value chains can identify influential traders via RUBNET. In the example from Indonesia, it appears that the most influential traders to address would be individuals who live within close physical proximity to many other stakeholders, belonging to the same ethnicity as the majority of the other stakeholders in their village, and are known to offer credit services to upstream suppliers. These traders could be identified by policy makers and used as information nodes. 




%
%
%
%
\clearpage 

\printbibliography


\newpage
\section{Appendix}
\hypertarget{appendix.ODD}{}

\subsection{Model description (ODD protocol)}

RUBNET is implemented in NetLogo version 6.0.3 \citep{Wilensky1999}. 
The model description follows the ODD (Overview, Design concepts, Details) protocol for describing individual-based models \citep{Grimm2006a,Grimm2010}.

\subsubsection{Overview}

\textbf{Purpose}\\
%
The main purpose of RUBNET is to understand channel choice behavior of rubber trading agents in Jambi, Sumatra, Indonesia: 
\emph{why do agricultural traders 
sell to a specific buyer?} \footnote{Parts of the \textquote{Overview} Section are included in the \textquote{Application} Section above.}

\textbf{Entities, state variables, and scales}\\
%
Main entities of RUBNET are the trading agents, represented by network nodes, and their potential interactions, represented by network links. Trading agents are grouped into two categories, selling agents (h0s) and buying agents (h1s). All selling agents are interconnected via social links representing the potential social interactions of these agents. Each selling agent is further connected to every buying agent via trading links, which represent potential trading connections (Figure \ref{model}). All these links are created during model initialization. However, not all links are considered active, which is indicated by a state variable of the links. Inactive links still exist in the model but are not shown in the visual output or considered for output measurements.
Each agent and link is characterized by a set of state variables (see Tables \ref{t.var.h0s}, \ref{t.var.h1s}, and \ref{t.var.links}). Most of the agent and link variables, such as locations, social characteristics, and trading information, are derived from empirical data. Thus, our model agents represent the original survey data distribution of agents and their characteristics.
RUBNET is spatially-explicit but does not utilize a discrete cell lattice with a specified grain and extent. Spatial distances between agents were calculated using geographical coordinates and are stored as a state variable of the links that connect the model agents. This is common practice for network models and allows to arrange agents and connections freely in space, without changing the spatial relations of the agents.

The selection of buying agents by the selling agents is performed by calculating scores for different criteria, such as distance or price. These criteria are weighted for score calculation. So-called \emph{global weights} applied for all selling agents are provided as global parameters (see Table \ref{t.var.global}). Additionally, each selling household has a set of individual weight preferences that are estimated from the selling agents' individual properties (see Table \ref{t.var.h0s}). These preferences are used to modify the global weights for final trading link score calculation.

\begin{table*}[h t b]
	\caption{State variables of selling agents \label{t.var.h0s}}
	\begin{tabular}[l]{p{4cm} p{10cm}}	
	\toprule
	state variable        & comments                                                 \\ \midrule
	h\_village            & village ID of the agents location                        \\
	h\_subdistrict        & subdistrict ID of the agents location                    \\
	h\_district           & district ID of the agents location \\
	h\_employees          & number of employees                                  \\
	h\_GPS\_S             & GPS southing of agents location                        \\
	h\_GPS\_E             & GPS northing of agents location                           \\
	h\_education 		  & Education level          \\
	h\_ethnicity          & Ethnicity                     \\ 
	h\_transport          & Transport capacity                       \\
	h\_prestigious\_job           & Important standing within social community                \\
    h\_activegroup        & Active in a group within the social community             \\
    h\_group\_count       & Number of groups the agent is active in    \\
    h\_hhsvlg             & Number of households in the same village \\
    h\_income             & Total income of selling agent   \\
    h\_age                & Age of selling agent             \\
    h\_hhsize             & Size of selling agents household  \\
    h\_housevalue         & House value of the selling agents house  \\
    h\_debt\_total\_mio\_log     & Amount of debts with specific trading agents       \\
    h\_n\_buyer           & Number of connected buying agents based on empirical data \\
    h\_w\_price           & Individual weight preference price criterion  \\
    h\_w\_dist           & Individual weight preference distance criterion  \\
    h\_w\_debts           & Individual weight preference debts criterion  \\
    h\_w\_social           & Individual weight preference social criterion  \\
	\bottomrule
\end{tabular}
\end{table*}

\begin{table*}[h t b]
	\caption{State variables of buying agents \label{t.var.h1s}}
	\begin{tabular}[l]{p{4cm} p{10cm}}	
		\toprule
		state variable        & comments                                    \\ \midrule
		h\_id				& Agent ID                                      \\
		h\_price            & Average buying price of the buying agent      \\
		\bottomrule
	\end{tabular}
\end{table*}

\begin{table*}[h t b]
	\caption{State variables of links \label{t.var.links}}
	\begin{tabular}[l]{p{4cm} p{10cm}}	
		\toprule
		state variable        & comments                                   \\ \midrule
		l\_length			& Real length of the connection, derived from GIS euclidean distance matrix    \\
		l\_tons            & Only trading links, amount of rubber traded via this link based on empirical data     \\
		l\_price            & Average buying price of this trading link, based on empirical data   \\
		l\_debts            & Amount of debts of selling agent with connected buying agent      \\
		l\_social            & Mean total score for buying agent from socially connected selling agents  \\
		l\_score           & trading links: final scores; social links: social influence  \\
        l\_score\_t1       & Temporarily stores the final score for the next model iteration \\
        l\_score\_price     & Sub-score price criterion \\
        l\_score\_dist     & Sub-score distance criterion \\
        l\_score\_debts     & Sub-score debts criterion \\
        l\_score\_social     & Sub-score social criterion \\
		l\_status\_model            & Indicates if the link is active (=chosen) in the model     \\
		l\_status\_data            & Indicates if the trading link exists within the empirical data     \\
		\bottomrule
	\end{tabular}
\end{table*}

\begin{table*}[h t b]
	\caption{Global parameters \label{t.var.global}}
	\begin{tabular}[l]{p{4cm} p{10cm}}	
		\toprule
		parameter        & comments                                   \\ \midrule
        n\_social          & Maximum number of social connections of each selling agent \\
		w\_price			& Weight of price criterion for trading selection    \\
		w\_dist            & Weight of distance criterion for trading selection     \\
		w\_debts            & Weight of debts criterion for trading selection   \\
		w\_social            & Weight of social interaction criterion for trading selection      \\
		w\_social\_education            & Weight of education level criterion for social interaction estimation     \\
		w\_social\_ethnicity           & Weight of ethnicity criterion for social interaction estimation    \\
		w\_social\_activegroup            & Weight of group activity criterion for social interaction estimation      \\
		w\_social\_prestigious\_job            & Weight of job influence criterion for social interaction estimation      \\
		w\_social\_proximity            & Weight of proximity criterion for social interaction estimation    \\
		\bottomrule
	\end{tabular}
\end{table*}

\textbf{Process overview and scheduling}\\
%
First, the model is initialized (see Figure \ref{f.flowchart}: for details on initialization, see Section \ref{sec_init}). Based on empirical data from an extensive survey of rubber traders, selling and buying model agents with specific agent IDs are created. Directed social links are created from each selling agent to each other selling agent and directed trading links are created from each selling agent to each buying agent. Geographical distances between agents are loaded from a distance matrix that provides Euclidean distances for all agent ID pairs. These distances are stored as link state variables. In order to initialize the trading related state variables of links and agents, the trading data set from the survey is loaded into the model. The data provides actual trading connections, including variables such as amount of rubber traded, selling and buying prices, debts between selling and buying agents, and social characteristics of agents.
After all agents and links are initialized, preliminary scores are calculated for each outgoing trading link of each selling agent. First, each selling agent calculates sub-scores for all outgoing trading links for the criteria price, distance, and debts. These individual scores are stored as an individual characteristic of the trading links. Then a preliminary final score is calculated using the global weights variables and the individual weight preferences.

The model execution covers four main procedures: determination of social interactions, calculation of final trading scores, selection of active trading connections, and calculation of output metrics (see Figure \ref{f.flowchart}). To determine social interactions, a social influence score is calculated for each social link, based on the properties of the two connected selling agents (for details see Section \ref{sec_submodel_social}). Depending on the parameter n\_social each selling agent selects the most influential incoming social links to be active. These active social links are considered the social network in terms of interactions between selling agents. Afterwards, final scores for each trading link are calculated. In contrast to the preliminary scores, these final scores also incorporate the trading decisions of socially connected selling agents. Finally, active trading connections are selected based on the final trading scores and output metrics of the emerging trading network are calculated.

\newpage
\label{BasicPrinciples}{~}
\subsubsection{Design concepts}

The ABS is designed to allow for testing of the hypotheses as laid out above.

\textbf{Basic principles}\\
%
Analogous to Figure \ref{FigureCausalChain}, \emph{three sets of variables} enter the simulation: first are \emph{stable determinants}, which include characteristics of buyer-seller-pairs that are constant over time. The second set are \emph{variable determinants (1)}. These include variable characteristics of buyer-seller-pairs. Set three consists of \emph{variable determinants (2)}, which represent the network effects. 

\textbf{Set one }includes the probability of interaction based on characteristics of the sellers and the buyers, such as the location in the same village (binary) or the bilateral distance (discrete, measured as the Euclidean distance). 
\textbf{The second set} consists of the size of a credit taken by the seller from one of the potential buyers. 
The CAS properties of \textbf{set three}, characterized by network effects, are captured by a so-called \emph{social matrix}. This matrix includes plausible channels via which the decision of seller A selling to buyer X increases the probability of seller B also selling to buyer X. The characteristics accounted for in the social matrix are the characteristics of the relationship between the sellers, such as the physical proximity (same as above), the similarity in the level of education, a shared ethnicity, the membership in any village group, or whether the potential influencer works in a prestigious job apart from the trading business. 


\newpage
\textbf{Emergence}\\
The resulting network of active trading connections emerges from the individual ranking decisions of selling agents. These decisions are influenced by the global weight parameters, individual weight preferences, and the trading connection decisions of socially connected selling agents.

\textbf{Adaption}\\
Trading decisions of model agents are influenced by other connected individuals based on the so-called \emph{social matrix}. Thus, agents in RUBNET adapt their behavior according to socially connected selling agents. RUBNET uses the social matrix parameters to rank social similarity of selling agents. Agents with similar social characteristics, such as home village, ethnicity, or education, have a higher chance to adapt trading decisions from each other, dependent upon the current social weights. The overall influence of this adaptive process on the final trading ranking of selling agents is defined by the global weight parameter w\_social. High values increase the probability that selling agents which share specific social characteristics use similar trading connections.






\textbf{Stochasticity}\\
Random decisions are only performed if a selling agent must decide between a set of buying agents with identical total scores. Apart from that, RUBNET is a deterministic model.


\textbf{Observation}\\
%
The main observation is the network structure that emerges from the individual trading connection decisions of selling agents. Several output metrics are derived from this active trading network (see Table \ref{t.observation}).

\begin{table*}[h t b]
	\caption{RUBNET observation metrics \label{t.observation}}
	\begin{tabular}[l]{p{4cm} p{10cm}}	
		\toprule
		variable       			& comments                                   \\ \midrule
		active\_tradings\_n		& Number of active trading links in the network    \\
		correct\_tradings\_n    & Number of trading links in the current network that match empirical trading links   \\
		correct\_tradings\_p    & Proportion of trading links in the current network that match empirical trading links  \\
		components\_n           & Number of network components of the active trading network  \\
		components\_size\_mu    & Mean component size of the active trading network    \\
		\bottomrule
	\end{tabular}
\end{table*}

\subsubsection{Details}

\textbf{Initialization}\\ \label{sec_init}
%
The initialization procedure contains five sub-procedures (see Figure \ref{f.flowchart}). (1) The survey input data file is loaded in order to create all selling and buying agents. These agents are initialized with the IDs from the survey data. Next, the input data is used to set all state variables (see Tables \ref{t.var.h0s} and \ref{t.var.h1s}) of the selling and buying agents. To initialize the networks, a directed social link is created from each selling agent to each other selling agent and a directed trading link is created from each selling agent to each buying agent. All these links are initialized with an inactive model and data status. (2) The model uses a distance matrix that contains Euclidean distances for each pair of agent IDs to store the distances between agents as a link state variable. (3) The survey input data is used to indicate which trading links have been observed in reality by setting the state variable \textit{l\_status\_data} to 1. (4) Individual weighting preferences are set up for each selling agent. These preferences are stored as properties of the selling agents and are used to modify the global weights for calculating final scores of trading links. By incorporating these individual weight preferences, selling agents may have different weighting priorities depending on their individual properties. 
\begin{itemize}
\item We assume a lower distance preference ($p_{di}$) if the transport capacity of the selling agent is high (1 / h\_transport)
\item We assume a lower price preference ($p_{p}$) if the selling agent is wealthy, indicated by the buying price of his/her house (1 / h\_housevalue)
\item We assume a lower debt preference ($p_{de}$) if the age of the selling agent is high (1 / h\_age)
\item We assume a lower social preference ($p_{soc}$) if the age of the selling agent is high (1 / h\_age)
\end{itemize}

These preferences are first calculated individually for each selling agent. Then, in a second step, all weights of each criterion are normalized using the absolute minimum and maximum values of each preference criterion. After normalization, all individual preferences have values between 1 and 2. For example, if the individual distance preference is 1, the global distance weight will be applied as it is. If the individual distance preference is 2, the global distance weight will be doubled, indicating a higher individual preference for the distance criterion.

(5) Initial scores are calculated for each outgoing trading connection of the selling agents. Initial scores are only calculated for the distance, price, and debts criteria, because distance to buying traders, prices offered by the buying traders, and debts incurred from a specific buying trader do not change during model execution. First, for each trading link a score is calculated for each criterion ($s_{p}$ for price, $s_{di}$ for distance, $s_{de}$ for debts). This is done by re-scaling the link values of each criterion into values between 0 and 1. For example, for the price criterion, the trading link with the highest price has a score of 1, whereas the trading link with the lowest price has a score of 0. For the distance criterion, reciprocal values are used to calculate scores, i.e. the longest trading link has a score of 0 and the shortest trading link has a score of 1.
Based on these criterion scores, each selling agent calculates a preliminary score for each outgoing trading link. For each initial weighting criterion a weighted score is calculated by multiplying the criterion score with its corresponding global weight ($w_p$ for price, $w_{di}$ for distance, $w_{de}$ for debts) and the selling agents' corresponding weight preferences (see Equation \ref{eq.initial.score}). The sum of these weighted scores is then divided by the sum of all weights, multiplied with the corresponding weight preferences.


\begin{equation}
\label{eq.initial.score}
l\_score = \frac{s_{p} * w_{p} * p_{p} + s_{di} * w_{di} * p_{di} + s_{de} * w_{de} * p_{de}}{w_{p} * p_{p} + w_{di} * p_{di} + w_{de} * p_{de}}
\end{equation}

\textbf{Input data}\\
This study is based on data generated with agricultural traders in the Jambi Province on Sumatra island, Indonesia, in 2012.\footnote{Data from subproject \href{http://www.uni-goettingen.de/en/c01+-+smallholder+productivity\%2c+market+access\%2c+and+international+linkages+in+rubber+and+palm+oil+production+in+jambi+province/412105.html}{C01} of \href{http://www.uni-goettingen.de/en/310995.html}{Collaborative Research Centre 990}. More information on the data is provided in \citet{Kopp2017a}.} In 40 randomly selected villages (stratified on a sub-district level), a total of 221 traders were interviewed, which is 71\% of all rubber traders active in the observed villages. The total population was determined by a snowball-like search. An overview of the variables entering the simulation is provided in Tables \ref{t.var.h0s} and \ref{t.var.h1s} and further descriptive statistics can be found in \hyperlink{appendix.tables}{appendix 6.2} (Tables \ref{t.descriptives.categorial} and \ref{t.descriptives.continuous}). 

As can be seen in Figure \ref{f.network}, some of the sample villages in close geographical proximity by while others are further apart. This means that sellers in sparse areas (in terms of villages per area) are over-sampled while the ones in dense areas are under-sampled. This leads to an over-representation in our sample of buyers who buy from only one seller. 
By construction, these links cannot be predicted by the influence of other sellers who already sell to these buyers, because there are none. This was solved by running the simulation with a sub-sample that excludes the buyers who buy from only one seller. As a robustness check the simulation was carried out with the unrestricted sample (Figure \ref{f.results.nullmodels}, bar \textquote{RUBNET complete}).

\textbf{Sub-models}
\vspace{-0.8cm}
%
\paragraph{Social interactions.} \label{sec_submodel_social}

It seems plausible that agents who are socially connected influence each other and tend to connect to similar traders. 
In order to allow for these interactions, each model loop starts with the determination of social interactions between selling agents. This is done by calculating a social score for each social link that depends on several properties of the selling agents that are connected by the social link. Social link scores always indicate the influence of selling agent A on selling agent B. To determine social scores, the properties of the selling agents are compared:

\begin{itemize}
\item The \emph{proximity} criterion has a value of 0 if agent A and B have different district IDs, a value of 0.33 if they have the same district ID, a value of 0.66 if they have the same district and subdistrict ID and a value of 1 if they have the same district, subdistrict and village ID.
\item The \emph{education} criterion has a value of 1 if agent A has the highest education (6) and a value of 0 if agent A has the lowest education (1). Values in between extremes are linearly decreasing depending on the education level (5:0.8, 4:0.6, 3:0.4, 2:0.2).
\item The \textquote{ethnicity} criterion has a value of 1 if agent A and B belong to the same ethnicity, and a value of 0 if they belong to different ethnicities.
\item The \emph{group activity} criterion has a value of 1 if agent A is active in 4 groups, and a value of 0 if agent A is active in 0 groups. Values in between extremes are linearly decreasing depending on the number of active groups (3:0.75, 2:0.5, 1:0.25).
\item The \emph{prestiguous job} criterion has a value of 1 if agent A has a prestigious job, and a value of 0 if agent A does not have a prestigious job.
\end{itemize}

The final social score of each social link is then calculated by weighting the social sub-scores with the global weight parameters for the social matrix (see Equation \ref{eq.social.score}). Since a social link is created from each selling agent to each other selling agent, two links exist between each pair of selling agents. From each pair of social links between the same agents only the social link with the higher social score is kept in the model.

\begin{equation}
\label{eq.social.score}
l\_score_{soc} = \frac{s_{prox} * w_{s,prox} + s_{educ} * w_{s,educ} + s_{eth} * w_{s,eth} + s_{group} * w_{s,group} + s_{job} * w_{s,job}}{w_{s,proximity} + w_{s,educ} + w_{s,eth} + w_{s,group} + w_{s,job}}
\end{equation}

Finally, depending on the parameter n\_social each selling agent selects the most influential incoming social links, i.e. the incoming social links with the highest social score, to be active. These active social links are considered the social network in terms of interactions between selling agents.

\paragraph{Final trading score calculation.} \label{sec_submodel_finalscore}

\begin{figure}[h t b]
	\protect\caption{Illustration of social interactions. \label{f.social.scores}}
	\begin{center}
	\includegraphics[width=100mm]{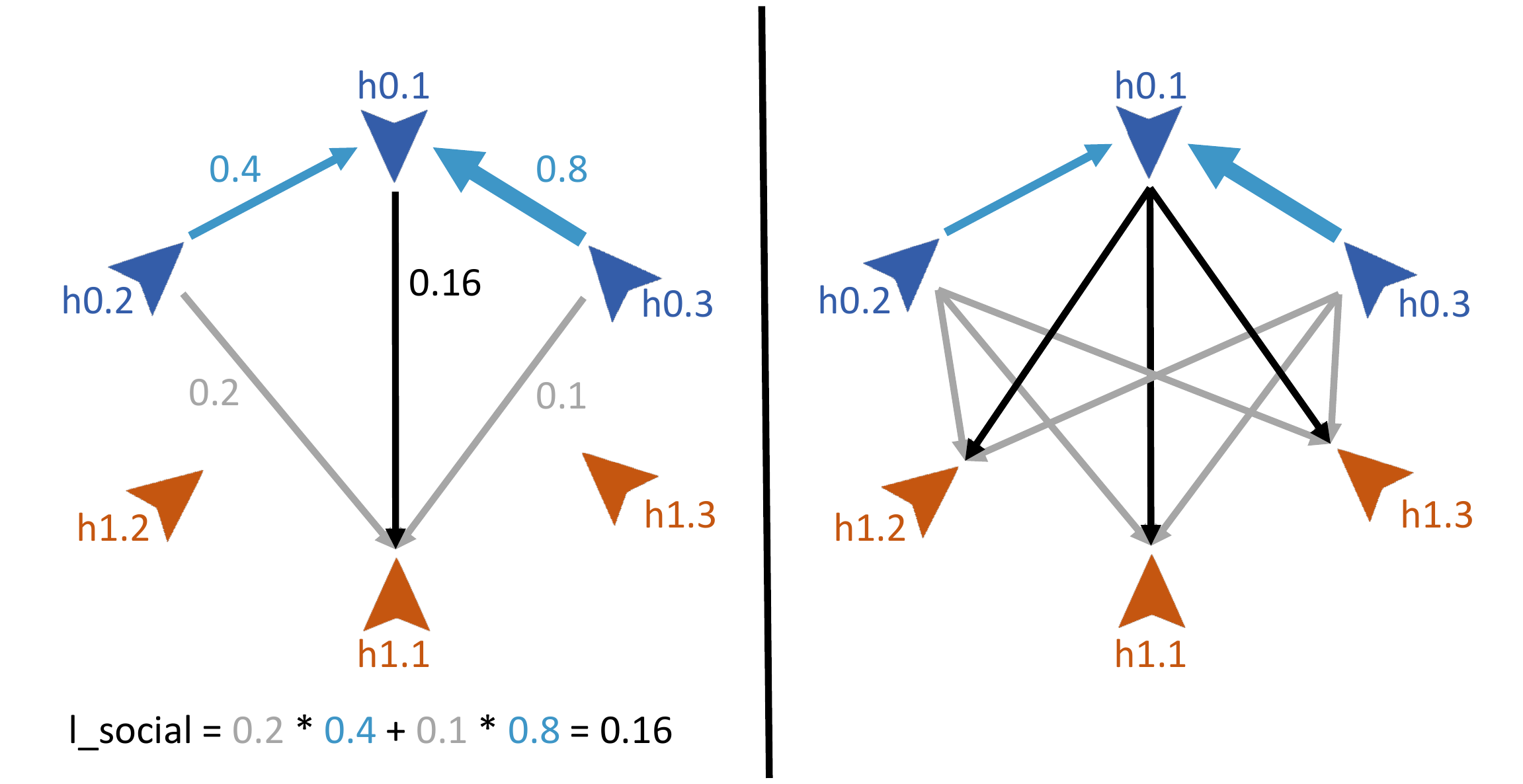}
	\end{center}
	\caption*{Illustration of social interactions. Blue triangles indicate selling agents (h0), while red triangles indicate buying agents (h1). Blue links indicate directed social links (social influence). To calculate the social score of the trading link from h0.1 to h1.1, the trading scores of socially connected sellers to that same buyer are determined (grey links). These scores are then weighted with the corresponding social influence (blue links) and summed up to calculate the social score of the trading link (black link). This is repeated for all buying links for each selling agent (right panel).}
\end{figure}

The trading link sub-scores for static criteria have been calculated during initialization (distance, price, debts). The social sub-scores of trading links are dynamic and are calculated during the procedure. This calculation is repeated in each loop iteration in the model, as trading choices may change in each iteration. Selling agents recommend buying agents to other selling agents within their social network. This is done by transferring the own individual trading scores of specific buyer agents to socially connected selling agents and weighting these scores with the strength of the social influence between these two selling agents (see Figure \ref{f.social.scores}).

For example, a selling agent h0.1 wants to determine the social score of the outgoing trading connection to buyer h1.1. First the \textquote{influencing} selling agents are determined, i.e. socially connected selling agents that have an influence on h0.1. Then the scores of the trading links of these socially connected selling agents to that same buyer (h1.1) are weighted by the social score and their sum is calculated.

This procedure is repeated for all selling agents and all outgoing trading connections. Afterwards, each trading connection has a social score value that represents the weighted score of socially connected selling agents for that same buyer. Finally, the social sub-scores of all trading links are normalized to an interval between 0 and 1.

The final trading score for each trading link is then calculated based on the initial sub-scores, which were calculated during initialization (distance, price, debts) and the social score indicating trading preferences within the selling agents' active social network (see Equation \ref{eq.final.score}).

\begin{equation}
\label{eq.final.score}
l\_score = \frac{s_{p} * w_{p} * p_{p} + s_{di} * w_{di} * p_{di} + s_{de} * w_{de} * p_{de} + s_{soc} * w_{soc} * p_{soc}}{w_{p} * p_{p} + w_{di} * p_{di} + w_{de} * p_{de} + w_{soc} * p_{soc}}
\end{equation}

\paragraph{Active trading selection.} \label{sec_submodel_choose}
Within this sub-model, each selling agent decides to which buying agent he or she will connect. The total number of active trading links a selling agents chooses is defined by the parameter \textit{h\_n\_buyer} which depends on the values of an agent's total sales. This relationship is found via a regression of the total sales value of each seller on the number of buyers that he/she is delivering to. The regression results can be found in Table \ref{RegressionNumberOfBuyers}. The number of links from each seller entering the simulation are the predicted values of this regression.
To select active trading connections, each selling agent simply selects the \textit{h\_n\_buyer} trading links with the highest final scores and sets the \textit{l\_status\_model} variables of these trading links to 1.



\paragraph{Output and measurements.} \label{sec_submodel_output}
Within this sub-model, all output observations are calculated (see Table \ref{t.observation}). Further, the visual output of RUBNET is updated. Inactive trading links are hidden and active trading links are displayed in different colors. These colors indicate whether a predicted active trading link has a matching trading connection within the empirically observed data.

\begin{figure}[!htbp]
	\protect\caption{Optimization process \label{f.results.optimization}}
\begin{center} 
	\includegraphics[width=0.6\columnwidth] {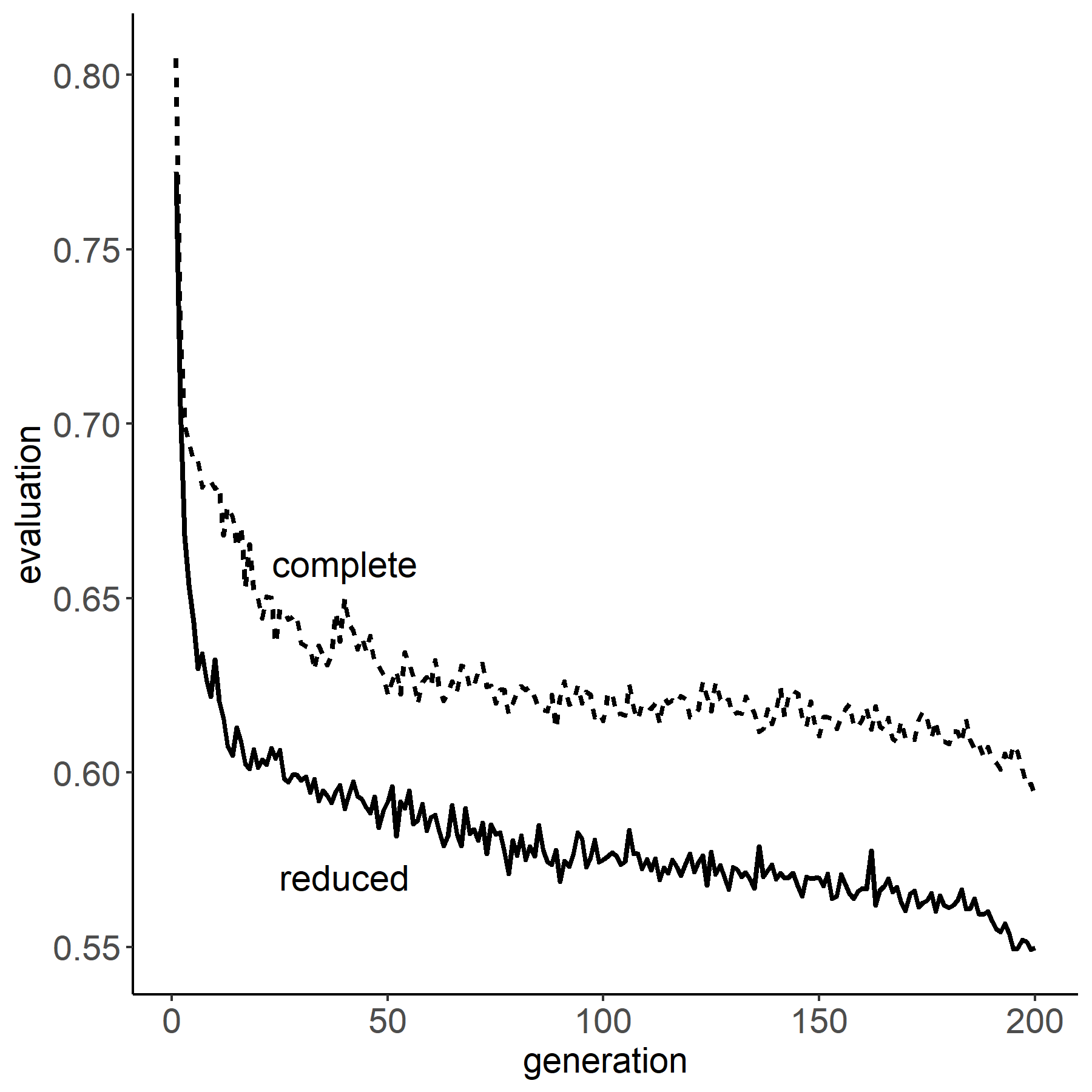}
\end{center}
\caption*{The share of wrongly predicted trading links is displayed on the y-axis. These decrease with repetitions of the optimization algorithm (x-axis).}
\end{figure}

\clearpage
\hypertarget{appendix.tables}{}
\subsection{Descriptive information on variables}

\begin{table*}[!htbp]
	\caption{Descriptives: categorial variables}
	\label{t.descriptives.categorial}
\begin{center}
	\begin{tabular}{lccc}
		\toprule
Ethnicity & Freq. & Percent & Description \\
		\midrule
1 & 71 & 39.66 & Javanese \\
2 & 94 & 52.51 & Jambinese \\
3 & 5 & 2.79 & Sundanese \\
4 & 3 & 1.68 & Melayan \\
5 & 6 & 3.35 & Other/unknown \\
		\midrule
Total & 179 & 100 & \\	
& & &		\\
Education & & &\\
\midrule
1 & 4 & 2.23 & Never went to school \\
2 & 19 & 10.61 & Attended primary school (not finished)\\
3 & 54 & 30.17 & Completed primary\\
4 & 33 & 18.44 & Completed secondary\\
5 & 4 & 2.23 & Completed post-secondary \\
6 & 65 & 36.31 & Completed college \\
		\midrule
Total & 179 & 100 & \\
%
& & &		\\
Prestigious job & & & \\
\midrule
& 168 & 94 & No prestigious job\\
& 11 & 6 &  Any kind of prestigious job\\
\midrule
1 & 2 & 18.18 & Village head  \\
5 & 1 & 9.09 & Village secretary \\
3 & 1 & 9.09 & Village council member \\
4 & 2 & 18.18 & Head of farmer group \\
2 & 1 & 9.09 & Chairman of youth organisation \\
6 & 4 & 36.36 & Teacher \\
		\midrule
Total & 179 & 100 &  \\
& & &		\\
Social engagement & & & \\
\midrule
0 & 109 & 60.89 & Not active in any village group \\
1 & 70$^*$ & 39.11 & Active in at least one village group \\
		\midrule
Total & 179 & 100 &  \\
		\midrule
\multicolumn{4}{c}{$^*$Includes members of farmers groups (33), neighbourhood groups (19),}\\
\multicolumn{4}{c}{religious groups (18), and 23 other groups (members in single digits).}\\
\end{tabular}
\end{center}
\end{table*}

\clearpage

\begin{table*}[!htbp]
	\caption{Descriptives: discrete and continuous variables}
	\label{t.descriptives.continuous}
\begin{center}
	\begin{tabular}{lcccc}
		\toprule
Variable & Mean & Std. Dev.  & Min & Max\\
\midrule
Number of employees & 5.7 & 6.8   & 0 & 50\\
Transport capacity/kg & 2060 & 2731.0  & 0& 16000 \\
\bottomrule
\end{tabular}
\end{center}
\end{table*}

\subsection{Relation between number of buyers and traded quantity}
\begin{table}[!htbp]
	\caption{Relation between number of buyers and traded quantity}
\label{RegressionNumberOfBuyers}
\begin{center}
\begin{tabular}{lc}
\toprule
 & (1) \\
VARIABLES & $number\_of\_buyers$ \\ 
\midrule
 &  \\
$\ln{total\_sales}$  & 0.065*** \\
 & (0.001) \\
Constant & 0.045** \\
 & (0.019) \\
 &  \\
Observations & 179 \\
 $R^2$ & 0.958 \\ 
\bottomrule
\multicolumn{2}{c}{ Standard errors in parentheses} \\
\multicolumn{2}{c}{ *** p$<$0.01, ** p$<$0.05, * p$<$0.1} \\
\end{tabular}
\caption*{Own production, based on survey data.}
\end{center}
\end{table}

\clearpage

\end{document}